# THE WIGNER FUNCTION NEGATIVE VALUE DOMAINS AND ENERGY FUNCTION POLES OF THE POLYNOMIAL OSCILLATOR

**E.E. Perepelkin[a,b,d], B.I. Sadovnikov[a], N.G. Inozemtseva[b,c], E.V. Burlakov[a,b], P.V. Afonin[a]**

[a] *Faculty of Physics, Lomonosov Moscow State University, Moscow, 119991 Russia*
[b] *Moscow Technical University of Communications and Informatics, Moscow, 123423 Russia*
[c] *Dubna State University, Moscow region, Moscow,141980 Russia*
[d] *Joint Institute for Nuclear Research, Moscow region, Moscow,141980 Russia*

**Abstract**

For a quantum oscillator with the polynomial potential an explicit expression that describes the energy distribution as a coordinate (and momentum) function is obtained. The presence of the energy function poles is shown for the quantum system in the domains where the Wigner function has negative values.

**Introduction**

With the development of quantum computing technique, quantum communication, cryptography and quantum information science, the mathematical apparatus of the Wigner function is becoming in demand. The Wigner function is used as a function of quasi-probabilities when describing a quantum system in a phase space [1-4]. A quantum feature of the Wigner function is the presence of negative values in the phase domain [5-8]. For example, for the implementation of many quantum cryptography schemes [9, 10] single photon sources must be used and this is one of the requirements that is necessary to protect the transmitted data on the most fundamental level [11]. However, creating a good source of single photons is a difficult enough challenge, while laser emission sources are widespread. In this regard, the idea arises [11] of generalizing the basic schemes of quantum cryptography to the case of nontrivial statistics of photons in the light modes of the transmitted signal. One of the main methods of investigating photon statistics is the homodyne detection method [12-14]. Applying this method to the signal under study, one may obtain experimental data on its so-called quadrature signal components, having which and using the methods of quantum tomography [15, 16], one may obtain the Wigner function of the signal under investigation. In this case, the Wigner function serves as a visual image of a quantum state, a kind of passport. The researcher may assess visually the «quality» of the quantum state he has obtained by comparing the obtained Wigner function with a certain standard. The presence of negative domains in the resulting Wigner function serves as a sufficient condition for the state to be «nonclassical».

For the simplest system, a quantum harmonic oscillator, the expression for the Wigner function is known explicitly:

$$W_s(\varepsilon) = \frac{(-1)^s}{\pi\hbar} e^{-2\varepsilon} L_s(4\varepsilon), \quad \varepsilon(x,p) = \frac{1}{\hbar\omega}\left(\frac{p^2}{2m} + \frac{m\omega^2 x^2}{2}\right), \qquad \text{(i.1)}$$

where $L_s$ are the Laguerre polynomials; $s$ is the number of the quantum state. Parameters $m, \omega$ stand for mass and frequency for the quantum harmonic oscillator. According to Hudson's theorem [17], only the Gaussian distribution of the Wigner function is positive. In the case of the quantum harmonic oscillator (i.1), it is the ground state $s=0$. In paper [18], for the quantum harmonic oscillator, the distributions of the average energy $\langle\tilde{\varepsilon}\rangle$ as a function of coordinates and momentum have been found explicitly:

$$\tilde{\varepsilon}(x,v) = 2\varepsilon(x,p) = \frac{v^2}{2\sigma_v^2} + \frac{x^2}{2\sigma_x^2},$$

$$\langle\tilde{\varepsilon}\rangle_{s,x}(x) \stackrel{det}{=} \frac{\int\limits_{-\infty}^{+\infty}\tilde{\varepsilon}(x,v)W_s(x,mv)dv}{\int\limits_{-\infty}^{+\infty}W_s(x,mv)dv}, \quad \langle\tilde{\varepsilon}\rangle_{s,v}(v) \stackrel{det}{=} \frac{\int\limits_{-\infty}^{+\infty}\tilde{\varepsilon}(x,v)W_s(x,mv)dx}{\int\limits_{-\infty}^{+\infty}W_s(x,mv)dx}, \quad (i.2)$$

$$\langle\tilde{\varepsilon}\rangle_{s,x}(x) = \frac{\langle v^2\rangle_{s,x}(x)}{2\sigma_v^2} + \frac{x^2}{2\sigma_x^2}, \qquad \langle\tilde{\varepsilon}\rangle_{s,v}(v) = \frac{v^2}{2\sigma_v^2} + \frac{\langle x^2\rangle_{s,v}(v)}{2\sigma_x^2}, \quad (i.3)$$

$$\langle v^2\rangle_{s,x}(x) = \sigma_v^2 \frac{\sum\limits_{k=0}^{s}C_k L_{s-k}\left(\frac{x^2}{\sigma_x^2}\right)}{\sum\limits_{k=0}^{s}\bar{C}_k L_{s-k}\left(\frac{x^2}{\sigma_x^2}\right)}, \qquad \langle x^2\rangle_{s,v}(v) = \sigma_x^2 \frac{\sum\limits_{k=0}^{s}C_k L_{s-k}\left(\frac{v^2}{\sigma_v^2}\right)}{\sum\limits_{k=0}^{s}\bar{C}_k L_{s-k}\left(\frac{v^2}{\sigma_v^2}\right)}, \quad (i.4)$$

$$C_k = (-1)^k \sum_{j=0}^{k} \frac{1+\eta(j)}{2} \frac{H_{k-j}^2(0) + 2(k-j)H_{k-j-1}^2(0)}{2^{k-j}(k-j)!},$$

$$\bar{C}_k = (-1)^k \sum_{j=0}^{k} \frac{1+\eta(j)}{2} \frac{H_{k-j}^2(0) - 2(k-j)H_{k-j-1}^2(0)}{2^{k-j}(k-j)!},$$

where $\sigma_x\sigma_v = \frac{\hbar}{2m}$, $\omega = \frac{\sigma_v}{\sigma_x}$; $H_k$ are the Hermite polynomials; $\eta(j)$ is the Heaviside step function. As shown in paper [18], the function of average values of energy $\langle\tilde{\varepsilon}\rangle_{s,v}$ and $\langle\tilde{\varepsilon}\rangle_{s,x}$ (i.3) have poles in the domains where corresponding Wigner functions $W_s(x,0)$ and $W_s(0,p)$ (i.1) have negative values of quasi-probability. In each domain of the Wigner function negativity there is one pole of the energy function. The number of the poles equals to the number of quantum state $s$.

The aim of this work is to extend the results obtained for a quantum harmonic oscillator [18] with quadratic potential $U_2 = \frac{m\omega^2 x^2}{2}$ to the general case of polynomial potential

$$U_N(x) = \sum_{n=0}^{N} a_n x^n, \quad (i.5)$$

where $a_n$ are known coefficients, for which there are «good» solutions of the Schrödinger equation. There is a number of reasons for us to consider the polynomial potential. First, any smooth potential can be approximated by potential (i.5) with the required accuracy. Thus, the result obtained for potential (i.5) can be used for a wide range of problems. Second, the Wigner function satisfies the Moyal equation [19] (a quantum analogue of the Liouville equation), which was obtained under the assumption that the potential of a quantum system is an analytical function, that is, it is represented as a formal power series of type (i.5). Choosing the value of $N$ in expression (i.5), it is possible with any accuracy to approximate the analytical function in the «oscillation point» neighborhood.

To achieve the set aim, we use the representation of the Wigner function in terms of the Weyl operator in the basis set of the harmonic oscillator [20]:

$$W(x,p) = \sum_{n,k=0}^{+\infty} \rho_{k,n} w_{n,k}(x,p) = \text{Tr}\left[\rho \mathcal{W}(x,p)\right], \tag{i.6}$$

$$w_{n,k}(x,p) = \frac{(-1)^n}{\pi \hbar} e^{-\kappa^2 x^2 - \frac{p^2}{\hbar^2 \kappa^2}} \mathcal{P}_{n,k}\left(-\kappa x - i\frac{p}{\hbar \kappa}, \kappa x - i\frac{p}{\hbar \kappa}\right),$$

$$\mathcal{P}_{n,k}(z_1, z_2) = \sqrt{2^{n+k} n! k!} \sum_{s=0}^{\min(n,k)} \frac{z_1^{n-s} z_2^{k-s}}{2^s s!(k-s)!(n-s)!},$$

where $\kappa = \sqrt{\frac{m\omega}{\hbar}}$; $\rho_{k,n} = c_k \bar{c}_n$ is a density matrix. Polynomials $\mathcal{P}_{n,k}$ admit the representation [20]

$$\mathcal{P}_{n,k}(-z, \bar{z}) = (-1)^n \Upsilon_{n,k}(|z|) e^{i(n-k)\varphi}, \tag{i.7}$$

$$\Upsilon_{n,k}(x) \stackrel{\text{det}}{=} x^{n+k} \sqrt{2^{n+k} n! k!} \sum_{s=0}^{\min(n,k)} \frac{(-1)^s}{2^s s!(k-s)!(n-s)! x^{2s}}.$$

that is

$$w_{n,k}(x,p) = \frac{1}{\pi \hbar} e^{-|z|^2} \Upsilon_{n,k}(|z|) e^{i(n-k)\varphi}, \tag{i.8}$$

$$|z|^2 = \frac{2}{\hbar \omega}\left(\frac{p^2}{2m} + \frac{m\omega^2 x^2}{2}\right) = 2\varepsilon(x,p), \qquad \varphi = \arg z = \text{arctg}\left(\frac{p}{m\omega x}\right).$$

Thus, using the representation for the Wigner function (i.6)-(i.8), one may obtain expressions for the average energies in an explicit form

$$\mathcal{E}_N(x,p) = \frac{p^2}{2m} + U_N(x), \tag{i.9}$$

$$\langle \mathcal{E}_N \rangle_{s,x}(x) = \frac{\int_{-\infty}^{+\infty} W_s(x,p) \mathcal{E}_N(x,p) dp}{\int_{-\infty}^{+\infty} W_s(x,p) dp}, \qquad \langle \mathcal{E}_N \rangle_{s,p}(p) = \frac{\int_{-\infty}^{+\infty} W_s(x,p) \mathcal{E}_N(x,p) dx}{\int_{-\infty}^{+\infty} W_s(x,p) dx}, \tag{i.10}$$

where $W_s(x,p)$ is the Wigner function of a quantum system with potential (i.5), corresponding to state $s$.

This work has the following structure. In §1, the theorems are proved on the form of explicit expressions for average kinematic values $\langle p^{2\ell} \rangle_{s,x}$ and $\langle x^r \rangle_{s,p}$, where $\ell, r \in \mathbb{N}_0 = \mathbb{N} \cup \{0\}$. Using the explicit expressions for the average kinematic values, expressions are constructed for average energies (i.10). In §2, the presence of poles of the energy functions (i.10) is shown in the domains, where Wigner function $W_s$ has negative values. This is based on the results of §1 and the method [21] for finding density matrix $\rho$ (i.6) for a quantum system with a polynomial potential. The Conclusions section contains the discussion of the results obtained. The Appendices contain proofs of the theorems.

## §1 Average kinematic values

Let us obtain expression (i.10) for the function of the average value of energy $\langle \mathcal{E}_N \rangle_{s,x}(x)$ of a quantum system with potential (i.5). According to (i.6), the Wigner function $W_s(x,p)$ of the $s$−th state has the form:

$$W_s(x,p) = \sum_{n,k=0}^{+\infty} \rho_{k,n}^{(s)} w_{n,k}(x,p). \tag{1.1}$$

Let us calculate integrals in the denominator and numerator of expression (i.10), respectively:

$$Q_s^{(1)}(x) = \int_{-\infty}^{+\infty} W_s(x,p) dp = \sum_{n,k=0}^{+\infty} \rho_{k,n}^{(s)} \int_{-\infty}^{+\infty} w_{n,k} dp = \sum_{n,k=0}^{+\infty} \rho_{k,n}^{(s)} I_{n,k}^0(x), \tag{1.2}$$

$$Q_s^{(2)}(x) = \int_{-\infty}^{+\infty} W_s(x,p) \mathcal{E}_N(x,p) dp = \frac{1}{2m} \sum_{n,k=0}^{+\infty} \rho_{k,n}^{(s)} \int_{-\infty}^{+\infty} p^2 w_{n,k} dp + U_N(x) Q_s^{(1)}(x) =$$
$$= \frac{1}{2m} \sum_{n,k=0}^{+\infty} \rho_{k,n}^{(s)} I_{n,k}^1(x) + U_N(x) Q_s^{(1)}(x), \tag{1.3}$$

where

$$I_{n,k}^\ell(x) \stackrel{\text{det}}{=} \int_{-\infty}^{+\infty} p^{2\ell} w_{n,k}(x,p) dp, \ \ell \in \mathbb{N}_0. \tag{1.4}$$

Using notations (1.2)-(1.4), expression $\langle \mathcal{E}_N \rangle_{s,x}$ (i.10) takes the form:

$$\langle \mathcal{E}_N \rangle_{s,x}(x) = \frac{1}{2m} \frac{\sum_{n,k=0}^{+\infty} \rho_{k,n}^{(s)} I_{n,k}^1(x)}{\sum_{n,k=0}^{+\infty} \rho_{k,n}^{(s)} I_{n,k}^0(x)} + U_N(x). \tag{1.5}$$

Thus, finding function $\langle \mathcal{E}_N \rangle_{s,x}$ is reduced to the calculation of integrals of type (1.4). We consider the following statements.

***Theorem 1*** *The expression for function (1.4) has the form:*

$$I_{n,k}^\ell(x) = \frac{(m\hbar\omega)^{\ell+\frac{1}{2}}}{\hbar\sqrt{\pi}} \begin{cases} \sqrt{\dfrac{2^{3\max(n,k)-\min(n,k)}}{n!k!}} |n-k| \sum_{\lambda=0}^{\min(n,k)} \sum_{s=0}^{[|n-k|/2]} \sum_{\mu=0}^{\min(n,k)-\lambda+s} C_{\min(n,k)-\lambda+s}^\mu \times \\ \times \dfrac{(-1)^{\lambda+s} \lambda! |2(\ell+\mu)-1|!!}{2^{\lambda+2s+\ell+\mu+1} s} C_k^\lambda C_n^\lambda C_{|n-k|-s-1}^{s-1} \bar{x}^{n+k-2(\lambda+\mu)} e^{-\bar{x}^2}, & \text{if } n \neq k, \\ \sum_{\lambda=0}^n \sum_{s=0}^\lambda \dfrac{(-1)^{\lambda+n}}{\lambda!} 2^{\lambda-\ell-s} C_n^\lambda C_\lambda^s |2(\ell+s)-1|!! \bar{x}^{2(\lambda-s)} e^{-\bar{x}^2}, & \text{if } n = k, \end{cases} \tag{1.6}$$

where $C_n^k$ is the number of combinations and $\left.\dfrac{1}{s}C_{|n-k|-(s+1)}^{s-1}\right|_{s=0} = \dfrac{1}{|n-k|}$; $\bar{x} = \kappa x$.

The proof of Theorem 1 is given in Appendix A.

In some cases, it is convenient to work with series containing orthogonal polynomials. Therefore, representation (1.6) can be expressed in terms of generalized Laguerre polynomials. We consider the lemma.

**Lemma** Matrix elements $w_{n,k}(x,p)$ admit the representation in terms of generalized Laguerre polynomials $L_{\min(n,k)}^{(|n-k|)}$:

$$w_{n,k}(x,p) = \dfrac{(-1)^{\min(n,k)}}{\pi\hbar}\sqrt{\dfrac{2^{|n-k|}\min(n,k)!}{\max(n,k)!}}e^{-|z|^2}|z|^{|n-k|}L_{\min(n,k)}^{(|n-k|)}\left(2|z|^2\right)e^{i(n-k)\varphi}, \qquad (1.7)$$

where values $|z|$ and $\varphi$ are determined by expressions (i.8).

The proof of the lemma is given in Appendix B.

Note that, in the particular case when $n = k$, expression (1.7) transforms into known expressions for the Wigner function of the harmonic oscillator (i.1). Using the statement of the lemma, we can represent integral (1.4) in terms of generalized Laguerre polynomials.

**Theorem 2** The expression for function (1.4) takes the form:

$$I_{n,k}^{\ell}(x) = A_{n,k}(m\hbar\omega)^{\ell+\frac{1}{2}}\begin{cases} 2^{|n-k|}|n-k|\displaystyle\sum_{\lambda=0}^{\min(n,k)}\sum_{s=0}^{[|n-k|/2]}\sum_{\mu=0}^{s}\dfrac{(-1)^s}{4^s s}C_{|n-k|-s-1}^{s-1}\times \\ \times C_s^{\mu}G_{\lambda}^{2(\ell+\mu)}\bar{x}^{|n-k|-2\mu}L_{\min(n,k)-\lambda}^{(|n-k|-1)}\left(2\bar{x}^2\right)e^{-\bar{x}^2}, \text{ if } n \neq k, \\ \sqrt{\pi}\displaystyle\sum_{\lambda=0}^{n}\sum_{s=0}^{\lambda}\dfrac{(-1)^{\lambda}}{\lambda!}2^{\lambda-\ell-s}C_n^{\lambda}C_{\lambda}^{s}|2(\ell+s)-1|!!\bar{x}^{2(\lambda-s)}e^{-\bar{x}^2}, \text{ if } n = k, \end{cases} \qquad (1.8)$$

where $A_{n,k} \stackrel{\text{det}}{=} \dfrac{(-1)^{\min(n,k)}}{\pi\hbar}\sqrt{\dfrac{2^{|n-k|}\min(n,k)!}{\max(n,k)!}}$; $C_n^k$ is the number of combinations; $\bar{x} = \kappa x$.

Coefficients $G_{\lambda}^{2\beta}$ are determined by the recurrence relation:

$$G_{\lambda}^{2\beta} = \left(\beta + \lambda - \dfrac{1}{2}\right)G_{\lambda}^{2(\beta-1)} - \lambda G_{\lambda-1}^{2(\beta-1)}, \ \beta \in \mathbb{N}, \qquad (1.9)$$

where

$$G_{\lambda}^{0} = (-1)^{\lambda}\dfrac{\sqrt{\pi}}{2^{\lambda+1}\lambda!}H_{\lambda}^{2}(0), \quad G_{\lambda}^{2} = (-1)^{\lambda}\dfrac{\sqrt{\pi}}{2}\left\{\dfrac{1}{2^{\lambda+1}\lambda!}H_{\lambda}^{2}(0) + \sum_{r=1}^{\lambda}\dfrac{H_{\lambda-r}^{2}(0)}{2^{\lambda-r}(\lambda-r)!}\right\}.$$

The proof of Theorem 2 is given in Appendix B.

Note that to find coefficients $G_\lambda^{2\beta}$ (1.9), it is convenient to use the formulas for the zeros of the Hermite polynomials $H_{2\lambda}^2(0) = \dfrac{(2\lambda)!(2\lambda)!}{\lambda!\lambda!}$, $H_{2\lambda+1}^2(0) = 0$.

Using formulas (1.6) and (1.8) in expression (1.6), one may obtain the distribution of average energy $\langle \mathcal{E}_N \rangle_{s,x}$. We obtain a similar expression for distribution $\langle \mathcal{E}_N \rangle_{s,p}$ (i.10). The denominator and numerator of the fraction (i.10) have the form, respectively:

$$R_s^{(1)}(p) = \sum_{n,k=0}^{+\infty} \rho_{k,n}^{(s)} \int_{-\infty}^{+\infty} w_{n,k}(x,p)dx = \sum_{n,k=0}^{+\infty} \rho_{k,n}^{(s)} J_{n,k}^0(p), \qquad (1.10)$$

$$R_s^{(2)}(p) = \frac{p^2}{2m} R_s^{(1)}(p) + \sum_{r=0}^{N} a_r \sum_{n,k=0}^{+\infty} \rho_{k,n}^{(s)} \int_{-\infty}^{+\infty} x^r w_{n,k}(x,p)dx = \frac{p^2}{2m} R_s^{(1)}(p) + \sum_{r=0}^{N} a_r \sum_{n,k=0}^{+\infty} \rho_{k,n}^{(s)} J_{n,k}^r(p).$$

Consequently,

$$\langle \mathcal{E}_N \rangle_{s,p}(p) = T(p) + \sum_{r=0}^{N} a_r \langle x^r \rangle_{s,p}(p), \qquad (1.11)$$

$$\langle x^r \rangle_{s,p}(p) = \dfrac{\sum\limits_{n,k=0}^{+\infty} \rho_{k,n}^{(s)} J_{n,k}^r(p)}{\sum\limits_{n,k=0}^{+\infty} \rho_{k,n}^{(s)} J_{n,k}^0(p)}, \qquad J_{n,k}^r(p) \stackrel{\text{det}}{=} \int_{-\infty}^{+\infty} x^r w_{n,k}(x,p)dx, \ r \in \mathbb{N}_0, \qquad (1.12)$$

where $T(p) \stackrel{\text{det}}{=} \dfrac{p^2}{2m}$. Let us formulate the following statement.

**Theorem 3** Let the density matrix satisfy the condition $\rho_{n,k} = \rho_{k,n}$, then functions $J_{n,k}^r(p)$ in the expression for the average value $\langle x^r \rangle_{s,p}$, $r \in \mathbb{N}_0$ (1.12) have the form:

when $n \neq k$:

$$J_{n,k}^r(p) = \begin{cases} 0, \text{ if } |n-k|+r = 2\nu+1, \\ \dfrac{|n-k|}{2\hbar\kappa^{r+1}} \sqrt{\dfrac{2^{3\max(n,k)-\min(n,k)}}{\pi n!k!}} \sum\limits_{\lambda=0}^{\min(n,k)} \sum\limits_{s=0}^{[|n-k|/2]} \sum\limits_{\mu=0}^{\min(n,k)+s-\lambda} \dfrac{(-1)^{\lambda+s} \lambda!}{2^{\lambda+s+\mu+\nu} s} |2(\nu+\mu-s)-1|! \times \\ \times C_{\min(n,k)+s-\lambda}^{\mu} C_k^{\lambda} C_n^{\lambda} C_{|n-k|-s-1}^{s-1} \bar{p}^{2(\min(n,k)+s-\lambda-\mu)} e^{-\bar{p}^2}, \text{ if } |n-k|+r = 2\nu. \end{cases} \qquad (1.13)$$

If $n = k$

$$J_{n,n}^r(p) = \begin{cases} 0, \text{ if } r = 2\ell+1, \\ \dfrac{2^n n!}{\kappa^{2\ell+1} \hbar \sqrt{\pi}} \sum\limits_{\lambda=0}^{n} \sum\limits_{\mu=0}^{n-\lambda} \dfrac{(-1)^\lambda |2(\ell+\mu)-1|!!}{2^{\lambda+\ell+\mu} \lambda! [(n-\lambda)!]^2} C_{n-\lambda}^{\mu} \bar{p}^{2(n-\lambda-\mu)} e^{-\bar{p}^2}, \text{ if } r = 2\ell, \end{cases} \qquad (1.14)$$

where $\bar{p} = \dfrac{p}{\sqrt{m\hbar\omega}}$; $\nu = 0,1,...$

The proof of Theorem 3 is given in Appendix C.

Substituting expressions (1.13) and (1.14) into formula (1.12), we can determine $\langle x^r \rangle_{s,p}$, and, consequently, the average energy (1.11) $\langle \mathcal{E}_N \rangle_{s,p}$.

## §2 Example of the polynomial potential

As an example, we consider potential of the third degree ($N = 3$) $U_3(x)$ with coefficients $\{a_r\} = \{0, 0, 2, -0.2\}$ (i.5). The degree of the potential anharmonicity is determined by coefficient $a_3$. Solving the stationary Schrödinger equation with potential $U_3(x)$ by the method set forth in [21], we obtain wave functions $\Psi_s(x)$, where $s$ is the number of the quantum state. Fig. 1 shows the probability density graphs $|\Psi_s(x)|^2$ for the first four states $s = 0, 1, 2, 3$. In the case of harmonic oscillator $U_2(x)$ distributions $|\Psi_s(x)|^2$ are the even functions. As seen from Fig. 1, the graphs of distributions $|\Psi_s(x)|^2$ have a nonsymmetrical form with respect to the origin of coordinates, which is due to the presence of anharmonicity in potential $U_3(x)$. In addition to the shift of the distributions to the right, an increase in the amplitude of the distributions is observed.

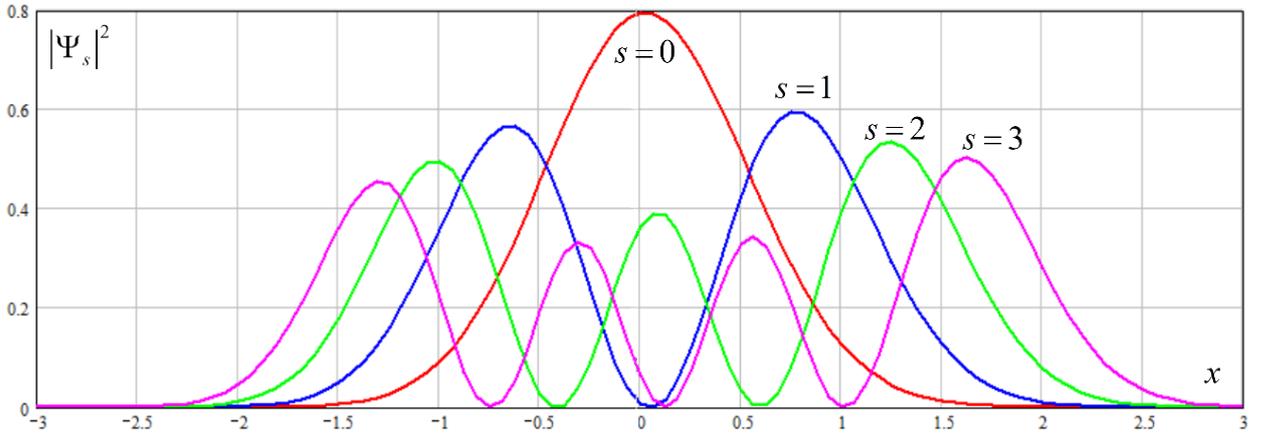

Fig. 1 Probability density $|\Psi_s|^2(x)$ for quantum states $s = 0, 1, 2, 3$

Fig. 2 shows the distributions of the Wigner functions $W_s(x, p)$ corresponding to wave functions $\Psi_s(x)$. A numerical-analytical method set forth in [21] was used for quick construction of function $W_s(x, p)$. In contrast to the Wigner functions of the harmonic oscillator, the distributions of quasi-density of probabilities $W_s(x, p)$ shown in Fig. 2 have explicitly nonsymmetrical form along coordinate axis $OX$ of the phase plane. The Wigner function of the ground state ($s = 0$) of the harmonic oscillator is a positive function (Hudson's theorem [17]). In Fig. 2, function $W_0(x, p)$ has a small domain of negative values, which is not noticeable in the Fig. 2. The presence of the negative values domain is directly related to the type of functions $w_{n,k}(x, p)$. It follows from representation (i.8) that function $w_{n,k}(x, p)$ in the phase plane has factorized form $R_{n,k}(\varepsilon)\Phi_{n,k}(\varphi)$. The angular part of $\Phi_{n,k}(\varphi)$ takes positive and negative values regardless of the values of $\varepsilon$ when $|n-k|\neq 0$. When $|n-k|=0$, only function $R_{n,n}(\varepsilon)$ is positive − $R_{0,0}(\varepsilon)$, that is the ground state of the harmonic oscillator (Gaussian distribution).

Distribution functions $W_s(x,p)$ resemble the Wigner functions of the harmonic oscillator, but due to the presence of anharmonicity in potential $U_3(x)$, they are not even in the coordinate.

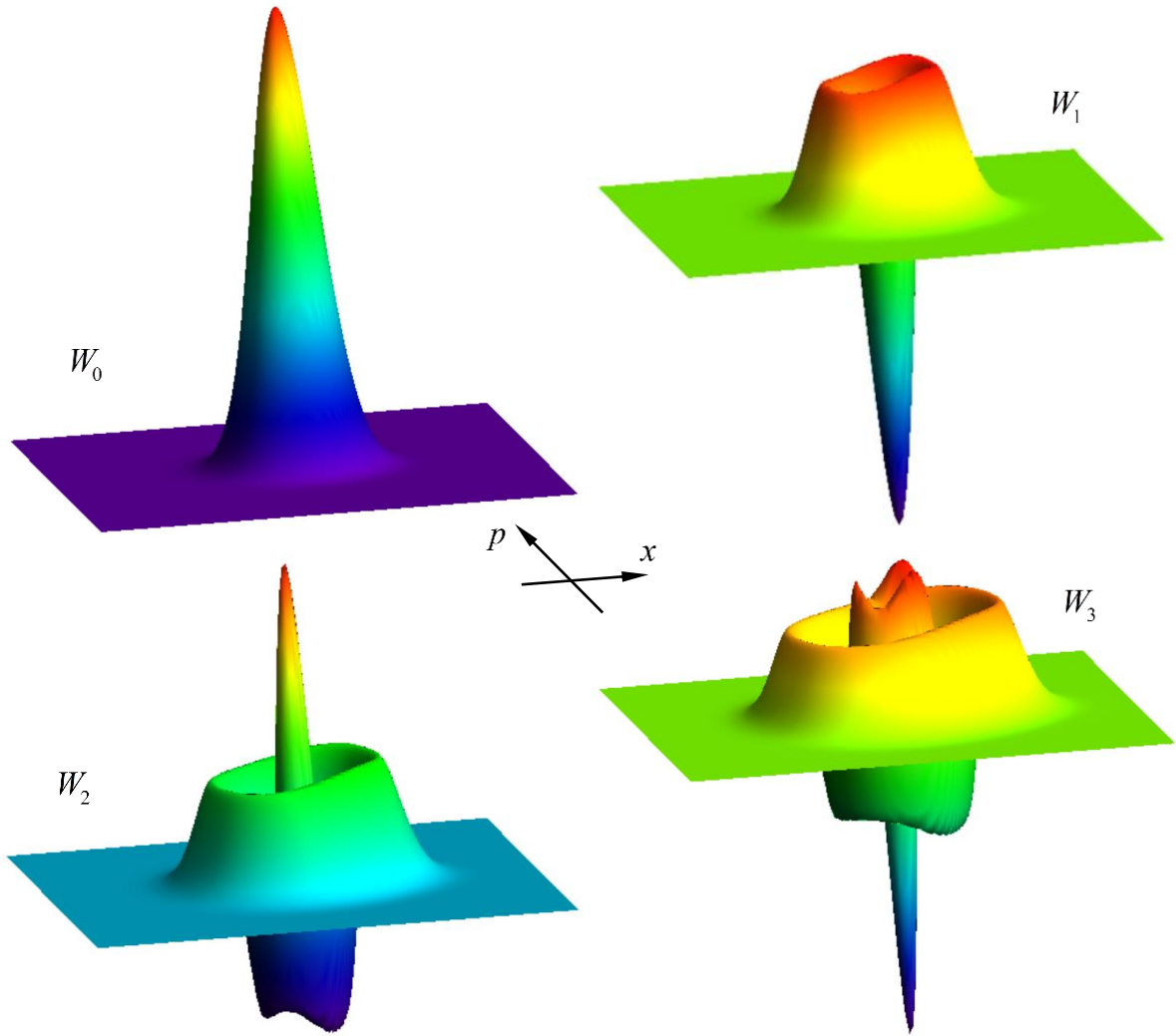

Fig. 2 Density of quasi-probabilities $W_s(x,p)$ for quantum states $s=0,1,2,3$

Fig. 3 shows the distributions of average energy $\langle\mathcal{E}_3\rangle_{s,x}$ and $\langle\mathcal{E}_3\rangle_{s,p}$ (blue) for three states $s=1,2,3$. The numerical calculations were performed using formulas (1.5) and (1.11). For the ground state ($s=0$), the graphs of energies $\langle\mathcal{E}_3\rangle_{0,x}$ and $\langle\mathcal{E}_3\rangle_{0,p}$ «almost» coincide with functions $U_3(x)$ and $T(p)$, respectively, therefore, they are absent in Fig. 3. In Fig. 3 on the left, distributions $\langle\mathcal{E}_3\rangle_{s,p}$ are shown along momentum axis $OP$, and on the right distributions $\langle\mathcal{E}_3\rangle_{s,x}$ are presented along coordinate axis $OX$. Distributions $U_3(x)$ and $T(p)$ in Fig. 3 are shown in red. Momentum distributions $\langle \mathcal{E}_3 \rangle_{s,p}$ are even functions, and distributions $\langle \mathcal{E}_3 \rangle_{s,x}$ have explicit nonsymmetrical form along the coordinate axis. All distributions $\langle \mathcal{E}_3 \rangle_{s,x}$ and $\langle \mathcal{E}_3 \rangle_{s,p}$ have poles, in which average energy $\langle \mathcal{E}_3 \rangle_s$ tends to infinity. The number of poles of the function of average energy $\langle \mathcal{E}_3 \rangle_s$ is determined by the number of quantum state $s$.

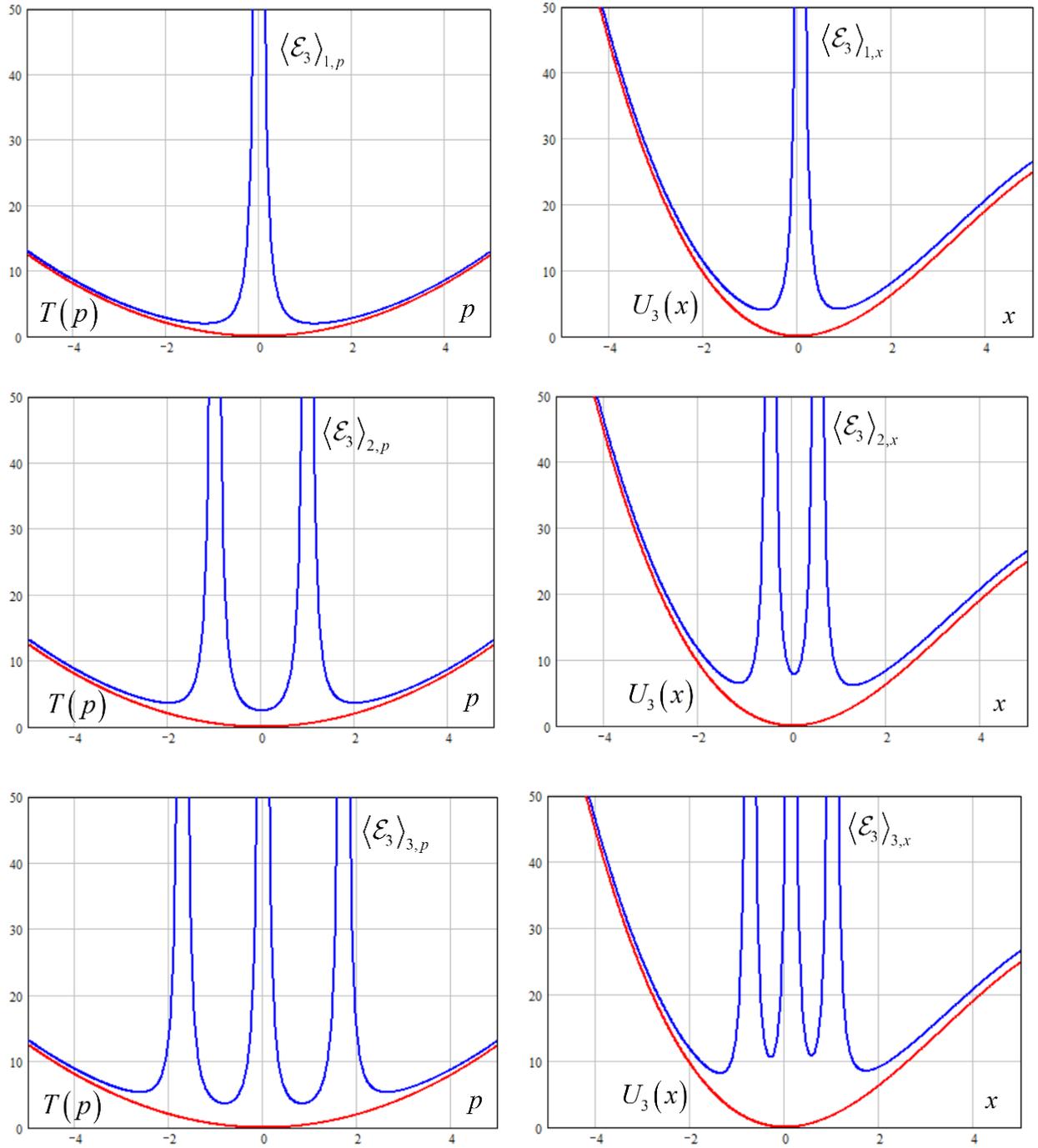

Fig. 3 Average energies $\langle \mathcal{E}_N \rangle_{s,p}$ and $\langle \mathcal{E}_N \rangle_{s,x}$ for quantum states $s = 1, 2, 3$

Let us compare the position of the poles of average energies $\langle \mathcal{E}_3 \rangle_{s,x}$ and $\langle \mathcal{E}_3 \rangle_{s,p}$ (see Fig. 3) to the domains where the corresponding distributions of the Wigner functions $W_s(x,0)$ and $W_s(0,p)$ (see Fig. 2) have the domains of negative values. Fig. 4 represents the superimposition of the data from Fig. 2 and Fig. 3. Distributions $W_s(x,0)$ and $W_s(0,p)$ are shown in Fig. 4 in blue. Black dots in Fig. 4 mark the position of the poles of average energies $\langle \mathcal{E}_3 \rangle_{s,x}$ and $\langle \mathcal{E}_3 \rangle_{s,p}$. It is seen from Fig. 4 that the poles of average energies $\langle \mathcal{E}_3 \rangle_{s,x}$ and $\langle \mathcal{E}_3 \rangle_{s,p}$ are located in the negative values domains of the distributions of the quasi-densities of probabilities $W_s(x,0)$ and $W_s(0,p)$, respectively.

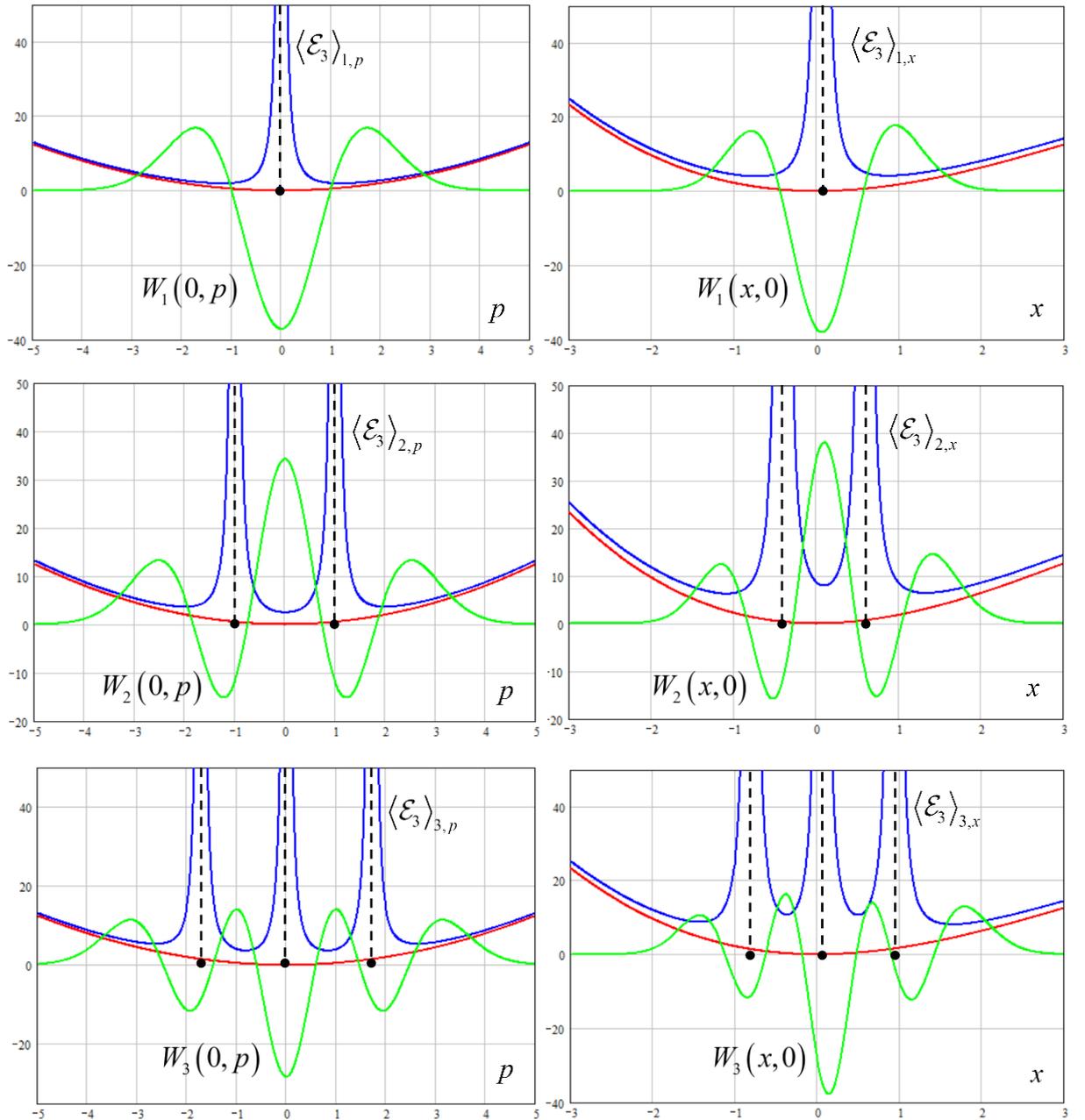

Fig. 4 Domains of negative values of the Wigner function and energy pole.

## Conclusions

As in the case of a harmonic oscillator (potential $U_2$ of the second degree), an oscillator with polynomial potential $U_N$ also has poles of the function of energy $\langle \mathcal{E}_N \rangle_{s,x}$ and $\langle \mathcal{E}_N \rangle_{s,p}$, located in the domains where the Wigner function takes negative values (see Fig. 4). The poles lead to the partition of the distribution function of energy $\langle \mathcal{E}_N \rangle_s$ into energy wells separated by infinitely high barriers (see Fig. 3). The number of such wells equals to the number of the oscillator quantum state $s$.

For a harmonic oscillator, the distributions $\langle \mathcal{E}_2 \rangle_{s,x}$ and $\langle \mathcal{E}_2 \rangle_{s,p}$ are symmetric due to the symmetry of the «energy» function $\varepsilon(x,p)$ (i.1) in the coordinate and momentum. For an oscillator with polynomial potential $U_N$, the symmetry of functions $\langle \mathcal{E}_N \rangle_{s,x}$ and $\langle \mathcal{E}_N \rangle_{s,p}$ is broken (see Fig. 3) by the presence of anharmonic terms in function $\mathcal{E}_N(x,p)$ (i.9).

Despite the fact that the ground state of an oscillator with a polynomial potential has a domain of negative values, the poles of functions $\langle \mathcal{E}_N \rangle_{0,x}$ and $\langle \mathcal{E}_N \rangle_{0,p}$ are absent. Note that the harmonic oscillator also has no poles for functions $\langle \mathcal{E}_2 \rangle_{0,x}$ and $\langle \mathcal{E}_2 \rangle_{0,p}$ of the ground state $(s=0)$.

The position of the poles of functions $\langle \mathcal{E}_N \rangle_{s,x}$ and $\langle \mathcal{E}_N \rangle_{s,p}$ is determined by the zeros of the distribution density functions of coordinates $|\Psi_s(x)|^2$ and momenta $|\tilde{\Psi}_s(p)|^2$, respectively (see Fig. 1 for the coordinate representation).

## Acknowledgements

This work was supported by the RFBR No. 18-29-10014. This research has been supported by the Interdisciplinary Scientific and Educational School of Moscow University «Photonic and Quantum Technologies. Digital Medicine».

## Appendix A

*Proof of Theorem 1*

Considering expression (i.8), integral (1.4) takes the form:

$$I_{n,k}^{\ell}(x) = \frac{1}{\pi\hbar} \int_{-\infty}^{+\infty} p^{2\ell} e^{-2\varepsilon} \Upsilon_{n,k}\left(\sqrt{2\varepsilon}\right) \cos(n-k)\varphi \, dp + \\ + \frac{i}{\pi\hbar} \int_{-\infty}^{+\infty} p^{2\ell} e^{-2\varepsilon} \Upsilon_{n,k}\left(\sqrt{2\varepsilon}\right) \sin(n-k)\varphi \, dp. \quad (A.1)$$

Integrating (A.1) over momentum $p \in (-\infty, +\infty)$, angle $\varphi \in \left(-\frac{\pi}{2}, \frac{\pi}{2}\right)$. Function $\sin\varphi = \sqrt{1 - \frac{\kappa^2 x^2}{2\varepsilon}} = \frac{\bar{p}}{\sqrt{2\varepsilon}}$, $\bar{p} = \frac{p}{\sqrt{m\hbar\omega}}$ is odd in variable $p$, consequently, function $\sin(n-k)\varphi$ will also be odd. As function $p^{2\ell} e^{-2\varepsilon} \Upsilon_{n,k}\left(\sqrt{2\varepsilon}\right)$ is even in variable $p$, then multiplying function by function $\sin(n-k)\varphi$ an odd function in variable $p$ is obtained. Therefore, the imaginary component of expression (A.1) equals to zero. As a result, expression (A.1) takes the form

$$I_{n,k}^{\ell} = \int_{-\infty}^{+\infty} p^{2\ell} \operatorname{Re}\left[w_{n,k}(x,p)\right] dp = \frac{1}{\pi\hbar} \int_{-\infty}^{+\infty} p^{2\ell} e^{-2\varepsilon} \Upsilon_{n,k}\left(\sqrt{2\varepsilon}\right) \cos(n-k)\varphi \, dp. \tag{A.2}$$

Expression $\cos\left[(n-k)\varphi\right]$ may be represented in terms of the Chebyshev polynomials

$$T_{n-k}(\cos\varphi) = \cos\left[(n-k)\varphi\right] = \cos\left[(k-n)\varphi\right] = T_{|n-k|}\left(\frac{\kappa x}{\sqrt{2\varepsilon}}\right), \tag{A.3}$$

where (i.8) is taken into consideration

$$\operatorname{tg}^2\varphi = \left(\frac{p}{m\omega x}\right)^2, \quad \cos^2\varphi = \frac{1}{1+\operatorname{tg}^2\varphi} = \frac{m^2\omega^2 x^2}{m^2\omega^2 x^2 + p^2},$$

$$\cos\varphi = \frac{m\omega x}{\sqrt{m^2\omega^2 x^2 + p^2}} = \frac{m\omega x}{\sqrt{2m\hbar\omega\varepsilon}} = \sqrt{\frac{m\omega}{\hbar}}\frac{x}{\sqrt{2\varepsilon}} = \frac{\kappa x}{\sqrt{2\varepsilon}},$$

$$p^2 + m^2\omega^2 x^2 = 2m\hbar\omega\varepsilon(x,p).$$

The Chebyshev polynomials admit representation in the form of a series:

$$T_n(x) = \frac{n}{2}\sum_{s=0}^{[n/2]}(-1)^s \frac{(n-s-1)!}{s!(n-2s)!}(2x)^{n-2s}, \quad n > 0,$$

or

$$T_{|n-k|}\left(\frac{\kappa x}{\sqrt{2\varepsilon}}\right) = \frac{|n-k|}{2}\sum_{s=0}^{[|n-k|/2]}(-1)^s \frac{(|n-k|-s-1)!}{s!(|n-k|-2s)!}\left(2\frac{\kappa x}{\sqrt{2\varepsilon}}\right)^{|n-k|-2s}. \tag{A.4}$$

Let us first consider the case when $n \neq k$. Substituting expressions (A.4) and (i.7)-(i.8) into integral (A.2), we obtain

$$\Upsilon_{n,k}\left(\sqrt{2\varepsilon}\right) = 2^{\frac{n+k}{2}}\varepsilon^{\frac{n+k}{2}}\sqrt{2^{n+k}n!k!}\sum_{\lambda=0}^{\min(n,k)}\frac{(-1)^\lambda}{2^{2\lambda}\lambda!(k-\lambda)!(n-\lambda)!\varepsilon^\lambda}, \tag{A.5}$$

$$I_{n,k}^{\ell} = \frac{1}{\pi\hbar}\int_{-\infty}^{+\infty} p^{2\ell} e^{-2\varepsilon} \Upsilon_{n,k}\left(\sqrt{2\varepsilon}\right) T_{|n-k|}\left(\frac{\kappa x}{\sqrt{2\varepsilon}}\right) dp =$$

$$= \frac{2^{n+k}\sqrt{n!k!}}{\pi\hbar}\sum_{\lambda=0}^{\min(n,k)}\frac{(-1)^\lambda}{2^{2\lambda}\lambda!(k-\lambda)!(n-\lambda)!}\int_{-\infty}^{+\infty} p^{2\ell} e^{-2\varepsilon}\varepsilon^{\frac{n+k}{2}-\lambda} T_{|n-k|}\left(\frac{\kappa x}{\sqrt{2\varepsilon}}\right) dp =$$

$$= \frac{2^{n+k+|n-k|-1}}{\pi\hbar}\sqrt{n!k!}\,|n-k|\sum_{\lambda=0}^{\min(n,k)}\sum_{s=0}^{[|n-k|/2]}\frac{(-1)^{\lambda+s}(|n-k|-s-1)!}{2^{2(\lambda+s)}\lambda!s!(k-\lambda)!(n-\lambda)!(|n-k|-2s)!} \times$$

$$\times \int_{-\infty}^{+\infty}\left(\frac{\kappa x}{\sqrt{2\varepsilon}}\right)^{|n-k|-2s} p^{2\ell} e^{-2\varepsilon}\varepsilon^{\frac{n+k}{2}-\lambda} dp = \frac{2^{2\max(n,k)-1}}{\pi\hbar}\frac{|n-k|}{\sqrt{n!k!}}\sum_{\lambda=0}^{\min(n,k)}\sum_{s=0}^{[|n-k|/2]}\frac{(-1)^{\lambda+s}\lambda!}{2^{2(\lambda+s)}}C_k^\lambda C_n^\lambda \frac{C_{|n-k|-s-1}^{s-1}}{s} \times$$

$$\times 2^{-\frac{|n-k|-2s}{2}} \overline{x}^{|n-k|-2s} \int_{-\infty}^{+\infty} \varepsilon^{\frac{n+k-|n-k|}{2}-\lambda+s} p^{2\ell} e^{-2\varepsilon} dp,$$

$$I_{n,k}^{\ell} = \frac{2^{2\max(n,k)-\min(n,k)-\frac{|n-k|}{2}}}{\pi\hbar} \frac{|n-k|}{\sqrt{n!k!}} (m\hbar\omega)^{\ell+\frac{1}{2}} \times$$
$$\times \sum_{\lambda=0}^{\min(n,k)} \sum_{s=0}^{[|n-k|/2]} \frac{(-1)^{\lambda+s} \lambda!}{2^{2(\lambda+s)-\lambda}} C_k^{\lambda} C_n^{\lambda} \frac{C_{|n-k|-s-1}^{s-1}}{s} \overline{x}^{|n-k|-2s} \int_0^{+\infty} (\overline{x}^2+\overline{p}^2)^{\min(n,k)-\lambda+s} \overline{p}^{2\ell} e^{-(\overline{x}^2+\overline{p}^2)} d\overline{p},$$
(A.6)

where the following taken into consideration

$$\frac{(|n-k|-s-1)!}{\lambda!s!(k-\lambda)!(n-\lambda)!(|n-k|-2s)!} = \frac{1}{\lambda!(k-\lambda)!(n-\lambda)!} \frac{(|n-k|-s-1)!}{s!(|n-k|-2s)!} =$$

$$= \frac{\lambda!}{k!n!} \frac{k!}{\lambda!(k-\lambda)!} \frac{n!}{\lambda!(n-\lambda)!} \frac{(|n-k|-(s+1))!}{s!(|n-k|-2s)!} = \frac{\lambda!}{k!n!} C_k^{\lambda} C_n^{\lambda} \frac{(|n-k|-(s+1))!}{s!(|n-k|-(s+1)-(s-1))!} =$$

$$= \frac{\lambda!}{k!n!} \frac{C_k^{\lambda} C_n^{\lambda}}{s} \frac{(|n-k|-(s+1))!}{(s-1)!(|n-k|-(s+1)-(s-1))!},$$

$$\frac{(|n-k|-s-1)!}{\lambda!s!(k-\lambda)!(n-\lambda)!(|n-k|-2s)!} = \frac{\lambda!}{k!n!} C_k^{\lambda} C_n^{\lambda} \frac{C_{|n-k|-s-1}^{s-1}}{s}.$$
(A.7)

Note that when $s=0$, there is no pole in expression (A.7) as

$$\left.\frac{1}{s} C_{|n-k|-(s+1)}^{s-1}\right|_{s=0} = \left.\frac{(|n-k|-s-1)!}{s!(|n-k|-2s)!}\right|_{s=0} = \frac{(|n-k|-1)!}{(|n-k|)!} = \frac{1}{|n-k|}.$$
(A.8)

Let us transform expression $2\max(n,k) - \min(n,k) - \frac{|n-k|}{2}$:

$$2\max(n,k) - \min(n,k) - \frac{|n-k|}{2} = \frac{3\max(n,k) - \min(n,k)}{2}.$$
(A.9)

Substituting (A.9) into expression (A.6) and use Newton's binomial formula:

$$I_{n,k}^{\ell} = \frac{\sqrt{2}^{3\max(n,k)-\min(n,k)}}{\pi\hbar} \frac{|n-k|}{\sqrt{n!k!}} (m\hbar\omega)^{\ell+\frac{1}{2}} \sum_{\lambda=0}^{\min(n,k)} \sum_{s=0}^{[|n-k|/2]} \frac{(-1)^{\lambda+s} \lambda!}{2^{\lambda+2s}} C_k^{\lambda} C_n^{\lambda} \frac{C_{|n-k|-s-1}^{s-1}}{s} \times$$
$$\times \sum_{\mu=0}^{\min(n,k)-\lambda+s} C_{\min(n,k)-\lambda+s}^{\mu} e^{-\overline{x}^2} \overline{x}^{2\min(n,k)+|n-k|-2(\lambda+\mu)} \int_0^{+\infty} \overline{p}^{2(\ell+\mu)} e^{-\overline{p}^2} d\overline{p},$$
(A.10)

taking into consideration that $2\min(n,k) + |n-k| = n+k$ and

$$\int_0^{+\infty} \overline{p}^{2(\ell+\mu)} e^{-\overline{p}^2} d\overline{p} = \frac{|2(\ell+\mu)-1|!!}{2^{\ell+\mu+1}} \sqrt{\pi},$$
(A.11)

expression (A.10) turns into (1.6) when $n \neq k$. If $n = k$, expression (i.1) can be used instead of expression (i.7)-(i.8), therefore

$$I_{n,n}^{\ell}(x) = \frac{(-1)^n}{\pi\hbar} \int_{-\infty}^{+\infty} p^{2\ell} e^{-2\varepsilon} L_n(4\varepsilon) dp = \frac{(-1)^n}{\pi\hbar} \sum_{\lambda=0}^{n} \frac{(-4)^{\lambda}}{\lambda!} C_n^{\lambda} \int_{-\infty}^{+\infty} p^{2\ell} e^{-2\varepsilon} \varepsilon^{\lambda} dp =$$

$$= 2\frac{(-1)^n}{\pi\hbar} \sum_{\lambda=0}^{n} \frac{(-4)^{\lambda}}{\lambda!} C_n^{\lambda} e^{-\frac{m\omega x^2}{\hbar}} \int_0^{+\infty} p^{2\ell} e^{-\frac{p^2}{m\hbar\omega}} \left(\frac{p^2}{2m\hbar\omega} + \frac{m\omega x^2}{2\hbar}\right)^{\lambda} dp =$$

$$= 2\frac{(-1)^n}{\pi\hbar} \sum_{\lambda=0}^{n} \frac{(-4)^{\lambda}}{\lambda!} C_n^{\lambda} e^{-\kappa^2 x^2} \sum_{s=0}^{\lambda} C_{\lambda}^s \left(\frac{\kappa^2 x^2}{2}\right)^{\lambda-s} \int_0^{+\infty} p^{2\ell} e^{-\frac{p^2}{m\hbar\omega}} \left(\frac{p^2}{2m\hbar\omega}\right)^s dp =$$

$$= 2\frac{(-1)^n}{\pi\hbar} (m\hbar\omega)^{\ell+\frac{1}{2}} \sum_{\lambda=0}^{n} \frac{(-2)^{\lambda}}{\lambda!} C_n^{\lambda} e^{-\kappa^2 x^2} \sum_{s=0}^{\lambda} C_{\lambda}^s \left(\kappa^2 x^2\right)^{\lambda-s} \int_0^{+\infty} \bar{p}^{2(\ell+s)} e^{-\bar{p}^2} d\bar{p},$$

$$I_{n,n}^{\ell}(x) = \frac{(m\hbar\omega)^{\ell+\frac{1}{2}}}{\hbar\sqrt{\pi}} \sum_{\lambda=0}^{n} \sum_{s=0}^{\lambda} \frac{(-1)^{\lambda+n}}{\lambda!} 2^{\lambda-\ell-s} C_n^{\lambda} C_{\lambda}^s |2(\ell+s)-1|!!(\kappa x)^{2(\lambda-s)} e^{-\kappa^2 x^2}. \qquad (A.12)$$

Theorem 1 is completely proved.

**Appendix B**

*Proof of Lemma*

Let us express polynomials $\Upsilon_{n,k}(x)$ in terms of generalized Laguerre polynomials. Without limiting the generality, we put $\ell = \min(n,k) = k$, $y = x\sqrt{2}$, $\lambda = k - s$ then

$$\Upsilon_{n,k}(x) = \left(\sqrt{2}x\right)^{n+k} \sqrt{n!k!} \sum_{s=0}^{\ell} \frac{(-1)^s}{s!(k-s)!(n-s)!\left(\sqrt{2}x\right)^{2s}} = y^{n+k} \sqrt{n!k!} \sum_{s=0}^{\ell} \frac{(-1)^s}{s!(k-s)!(n-s)!y^{2s}},$$

$$\Upsilon_{n,k}(x) = (-1)^k y^{n-k} \sqrt{n!k!} \sum_{\lambda=0}^{\ell} \frac{(-1)^{\lambda} y^{2\lambda}}{\lambda!(k-\lambda)!(n-k+\lambda)!} = y^{n-k} \frac{(-1)^k \sqrt{n!k!}}{(n-k+k)!} \sum_{\lambda=0}^{\ell} \frac{(-1)^{\lambda}(k+n-k)! y^{2\lambda}}{\lambda!(k-\lambda)!(n-k+\lambda)!},$$

Using substitution $\alpha = n - k$, we obtain

$$\Upsilon_{n,k}(x) = (-1)^k y^{\alpha} \sqrt{\frac{k!}{n!}} \sum_{\lambda=0}^{\ell} \frac{(-1)^{\lambda}(k+\alpha)! y^{2\lambda}}{\lambda!(k-\lambda)!(\alpha+\lambda)!} = (-1)^k y^{n-k} \sqrt{\frac{k!}{n!}} L_{\ell}^{(n-k)}(y^2),$$

$$\Upsilon_{n,k}(x) = (-1)^k x^{n-k} \sqrt{\frac{2^{n-k} k!}{n!}} L_{\min(n,k)}^{(n-k)}(2x^2), \qquad (B.1)$$

where it is taken into consideration that

$$L_{\ell}^{(\alpha)}(x) = \sum_{\lambda=0}^{\ell} \frac{(-1)^{\lambda}(\ell+\alpha)! x^{\lambda}}{\lambda!(\ell-\lambda)!(\alpha+\lambda)!}. \qquad (B.2)$$

If $\ell = \min(n,k) = n$, then $\lambda = n - s$

$$\Upsilon_{n,k}(x) = \left(\sqrt{2}x\right)^{n+k} \sqrt{n!k!} \sum_{s=0}^{\ell} \frac{(-1)^s}{s!(k-s)!(n-s)!\left(\sqrt{2}x\right)^{2s}} = y^{n+k} \sqrt{n!k!} \sum_{s=0}^{\ell} \frac{(-1)^s}{s!(k-s)!(n-s)!y^{2s}},$$

$$\Upsilon_{n,k}(x) = (-1)^n y^{k-n} \sqrt{n!k!} \sum_{\lambda=0}^{\ell} \frac{(-1)^\lambda y^{2\lambda}}{\lambda!(n-\lambda)!(k-n+\lambda)!} = y^{k-n} \frac{(-1)^n \sqrt{n!k!}}{(k-n+n)!} \sum_{\lambda=0}^{\ell} \frac{(-1)^\lambda (k-n+n)! y^{2\lambda}}{\lambda!(n-\lambda)!(k-n+\lambda)!},$$

$$\Upsilon_{n,k}(x) = (-1)^n y^{-\alpha} \sqrt{\frac{n!}{k!}} \sum_{\lambda=0}^{\ell} \frac{(-1)^\lambda (-\alpha+n)! y^{2\lambda}}{\lambda!(n-\lambda)!(-\alpha+\lambda)!} = (-1)^n y^{-\alpha} \sqrt{\frac{n!}{k!}} L_{\min(n,k)}^{(-\alpha)}(y^2),$$

$$\Upsilon_{n,k}(x) = (-1)^n x^{k-n} \sqrt{\frac{2^{k-n} n!}{k!}} L_{\min(n,k)}^{(k-n)}(2x^2). \tag{B.3}$$

Generalizing expression (B.1) and (B.3), we obtain finally

$$\Upsilon_{n,k}(x) = (-1)^{\min(n,k)} x^{|n-k|} \sqrt{\frac{2^{|n-k|} \min(n,k)!}{\max(n,k)!}} L_{\min(n,k)}^{(|n-k|)}(2x^2). \tag{B.4}$$

Considering representations (B.4) and (i.8), expression (i.7) for elements $w_{n,k}$ takes the form (1.7). The lemma is proved.

*Proof of Theorem 2*
Substituting expression (1.7) into integral (A.2), we obtain

$$I_{n,k}^\ell = A_{n,k} \int_{-\infty}^{+\infty} p^{2\ell} e^{-2\varepsilon} (2\varepsilon)^{\frac{|n-k|}{2}} L_{\min(n,k)}^{(|n-k|)}(4\varepsilon) \cos\left[(n-k)\varphi\right] dp. \tag{B.5}$$

Using the expression for the Chebyshev polynomials (A.4) and the expansion for generalized Laguerre polynomials:

$$L_n^{(\alpha+\beta+1)}(x+y) = \sum_{m=0}^n L_m^{(\alpha)}(x) L_{n-m}^{(\beta)}(y),$$

$$L_{\min(n,k)}^{(|n-k|)}\left(\frac{2p^2}{m\hbar\omega} + \frac{2m\omega x^2}{\hbar}\right) = \sum_{\lambda=0}^{\min(n,k)} L_\lambda^{(0)}\left(\frac{2p^2}{m\hbar\omega}\right) L_{\min(n,k)-\lambda}^{(|n-k|-1)}\left(\frac{2m\omega x^2}{\hbar}\right), \tag{B.6}$$

we obtain

$$I_{n,k}^{\ell} = A_{n,k} e^{-\frac{m\omega x^2}{\hbar}} \int_{-\infty}^{+\infty} p^{2\ell} e^{-\frac{p^2}{m\hbar\omega}} \sqrt{2\varepsilon}^{|n-k|} L_{\min(n,k)}^{(|n-k|)}(4\varepsilon) T_{|n-k|}\left(\kappa \frac{x}{\sqrt{2\varepsilon}}\right) dp =$$

$$= A_{n,k} e^{-\frac{m\omega x^2}{\hbar}} \frac{|n-k|}{2} \sum_{\lambda=0}^{\min(n,k)} L_{\min(n,k)-\lambda}^{(|n-k|-1)}\left(\frac{2m\omega x^2}{\hbar}\right) \times$$

$$\times \int_{-\infty}^{+\infty} p^{2\ell} e^{-\frac{p^2}{m\hbar\omega}} \sqrt{2\varepsilon}^{|n-k|} L_\lambda\left(\frac{2p^2}{m\hbar\omega}\right) \sum_{s=0}^{[|n-k|/2]} (-1)^s \frac{(|n-k|-s-1)!}{s!(|n-k|-2s)!} x^{|n-k|-2s} (2\kappa)^{|n-k|-2s} \sqrt{2\varepsilon}^{2s-|n-k|} dp =$$

$$= A_{n,k} e^{-\frac{m\omega x^2}{\hbar}} \frac{|n-k|}{2} (2\kappa)^{|n-k|} \sum_{\lambda=0}^{\min(n,k)} L_{\min(n,k)-\lambda}^{(|n-k|-1)}\left(\frac{2m\omega x^2}{\hbar}\right) \sum_{s=0}^{[|n-k|/2]} \left(\frac{-1}{4\kappa^2}\right)^s \frac{(|n-k|-s-1)!}{s!(|n-k|-2s)!} x^{|n-k|-2s} \times$$

$$\times \int_{-\infty}^{+\infty} p^{2\ell} e^{-\frac{p^2}{m\hbar\omega}} L_\lambda\left(\frac{2p^2}{m\hbar\omega}\right) \sqrt{2\varepsilon}^{2s} dp,$$

$$I_{n,k}^{\ell} = A_{n,k} e^{-\frac{m\omega x^2}{\hbar}} \frac{|n-k|}{2} (2\kappa)^{|n-k|} \sum_{\lambda=0}^{\min(n,k)} L_{\min(n,k)-\lambda}^{(|n-k|-1)}\left(\frac{2m\omega x^2}{\hbar}\right) \times$$

$$\times \sum_{s=0}^{[|n-k|/2]} \left(\frac{-1}{4\kappa^2}\right)^s \frac{(|n-k|-s-1)!}{s!(|n-k|-2s)!} x^{|n-k|-2s} J_\lambda^{\ell,s}(x), \quad (B.7)$$

$$J_\lambda^{\ell,s}(x) \stackrel{\text{det}}{=} \int_{-\infty}^{+\infty} p^{2\ell} e^{-\frac{p^2}{m\hbar\omega}} L_\lambda\left(\frac{2p^2}{m\hbar\omega}\right) \left(\frac{p^2}{m\hbar\omega} + \frac{m\omega x^2}{\hbar}\right)^s dp.$$

Let us calculate integral $J_\lambda^{\ell,s}(x)$. Let us introduce the notations $\bar{p} = \frac{p}{\sqrt{m\hbar\omega}}$, $\bar{x} = \kappa x$, we obtain

$$J_\lambda^{\ell,s}(x) = (m\hbar\omega)^{\ell+\frac{1}{2}} \int_{-\infty}^{+\infty} \bar{p}^{2\ell} e^{-\bar{p}^2} L_\lambda(2\bar{p}^2)(\bar{p}^2 + \bar{x}^2)^s d\bar{p} =$$

$$= (m\hbar\omega)^{\ell+\frac{1}{2}} \sum_{\mu=0}^{s} C_s^\mu \bar{x}^{2(s-\mu)} \int_{-\infty}^{+\infty} \bar{p}^{2\ell} e^{-\bar{p}^2} L_\lambda(2\bar{p}^2) \bar{p}^{2\mu} d\bar{p},$$

$$J_\lambda^{\ell,s}(x) = 2(m\hbar\omega)^{\ell+\frac{1}{2}} \sum_{\mu=0}^{s} C_s^\mu \bar{x}^{2(s-\mu)} \int_{0}^{+\infty} \bar{p}^{2(\ell+\mu)} e^{-\bar{p}^2} L_\lambda(2\bar{p}^2) d\bar{p}. \quad (B.8)$$

Let us calculate the integral from expression (B.8). Performing integration by parts, we obtain

$$\int_{0}^{+\infty} \bar{p}^{2(\ell+\mu)} e^{-\bar{p}^2} L_\lambda(2\bar{p}^2) d\bar{p} = \frac{1}{2(\ell+\mu)+1} \int_{0}^{+\infty} e^{-\bar{p}^2} L_\lambda(2\bar{p}^2) d\bar{p}^{2(\ell+\mu)+1} =$$

$$= -\frac{1}{2(\ell+\mu)+1} \int_{0}^{+\infty} \bar{p}^{2(\ell+\mu)+1} \frac{d}{d\bar{p}}\left[e^{-\bar{p}^2} L_\lambda(2\bar{p}^2)\right] d\bar{p}. \quad (B.9)$$

Using expression $L_\lambda'(x) = \frac{\lambda}{x}[L_\lambda(x) - L_{\lambda-1}(x)]$

$$\frac{d}{d\bar{p}}\left[e^{-\bar{p}^2}L_\lambda\left(2\bar{p}^2\right)\right]=-2\bar{p}e^{-\bar{p}^2}L_\lambda\left(2\bar{p}^2\right)+e^{-\bar{p}^2}L_\lambda{'}\left(2\bar{p}^2\right)4\bar{p}=$$

$$=2e^{-\bar{p}^2}\left[2\bar{p}L_\lambda{'}\left(2\bar{p}^2\right)-\bar{p}L_\lambda\left(2\bar{p}^2\right)\right]=2e^{-\bar{p}^2}\left[\frac{\lambda}{\bar{p}}\left(L_\lambda\left(2\bar{p}^2\right)-L_{\lambda-1}\left(2\bar{p}^2\right)\right)-\bar{p}L_\lambda\left(2\bar{p}^2\right)\right],$$

$$\frac{d}{d\bar{p}}\left[e^{-\bar{p}^2}L_\lambda\left(2\bar{p}^2\right)\right]=\frac{2}{\bar{p}}e^{-\bar{p}^2}\left[\left(\lambda-\bar{p}^2\right)L_\lambda\left(2\bar{p}^2\right)-\lambda L_{\lambda-1}\left(2\bar{p}^2\right)\right]. \tag{B.10}$$

Substituting expression for derivative (B.10) into integral (B.9), we obtain

$$\int_0^{+\infty}\bar{p}^{2(\ell+\mu)}e^{-\bar{p}^2}L_\lambda\left(2\bar{p}^2\right)d\bar{p}=-\frac{2}{2(\ell+\mu)+1}\int_0^{+\infty}\bar{p}^{2(\ell+\mu)}e^{-\bar{p}^2}\left[\left(\lambda-\bar{p}^2\right)L_\lambda\left(2\bar{p}^2\right)-\lambda L_{\lambda-1}\left(2\bar{p}^2\right)\right]d\bar{p}=$$

$$=-\frac{2\lambda}{2(\ell+\mu)+1}\int_0^{+\infty}\bar{p}^{2(\ell+\mu)}e^{-\bar{p}^2}L_\lambda\left(2\bar{p}^2\right)d\bar{p}+\frac{2}{2(\ell+\mu)+1}\int_0^{+\infty}\bar{p}^{2(\ell+\mu)+2}e^{-\bar{p}^2}L_\lambda\left(2\bar{p}^2\right)d\bar{p}+$$

$$+\frac{2\lambda}{2(\ell+\mu)+1}\int_0^{+\infty}\bar{p}^{2(\ell+\mu)}e^{-\bar{p}^2}L_{\lambda-1}\left(2\bar{p}^2\right)d\bar{p},$$

from this

$$\int_0^{+\infty}\bar{p}^{2(\ell+\mu)+2}e^{-\bar{p}^2}L_\lambda\left(2\bar{p}^2\right)d\bar{p}=$$

$$=\left(\ell+\mu+\lambda+\frac{1}{2}\right)\int_0^{+\infty}\bar{p}^{2(\ell+\mu)}e^{-\bar{p}^2}L_\lambda\left(2\bar{p}^2\right)d\bar{p}-\lambda\int_0^{+\infty}\bar{p}^{2(\ell+\mu)}e^{-\bar{p}^2}L_{\lambda-1}\left(2\bar{p}^2\right)d\bar{p},$$

or

$$G_\lambda^{2(\ell+\mu)+2}=\left(\ell+\mu+\lambda+\frac{1}{2}\right)G_\lambda^{2(\ell+\mu)}-\lambda G_{\lambda-1}^{2(\ell+\mu)}, \tag{B.11}$$

where

$$G_\lambda^{2\beta}=\int_0^{+\infty}\bar{p}^{2\beta}e^{-\bar{p}^2}L_\lambda\left(2\bar{p}^2\right)d\bar{p}. \tag{B.12}$$

In [18], the integrals were calculated:

$$\int_{-\infty}^{+\infty}e^{-\tau^2}L_k\left(2\tau^2\right)d\tau=(-1)^k\frac{\sqrt{\pi}}{2^k k!}H_k^2(0), \tag{B.13}$$

$$\int_{-\infty}^{+\infty}\tau^2 e^{-\tau^2}L_k\left(2\tau^2\right)d\tau=(-1)^k\sqrt{\pi}\left\{\frac{1}{2^{k+1}k!}H_k^2(0)+\sum_{s=1}^k\frac{H_{k-s}^2(0)}{2^{k-s}(k-s)!}\right\}.$$

Considering the result (B.13), integral (B.12) at $\beta=0$ and $\beta=1$ will take the from, respectively:

$$G_\lambda^0=(-1)^\lambda\frac{\sqrt{\pi}}{2^{\lambda+1}\lambda!}H_\lambda^2(0), \tag{B.14}$$

$$G_\lambda^2 = (-1)^\lambda \frac{\sqrt{\pi}}{2} \left\{ \frac{1}{2^{\lambda+1} \lambda!} H_\lambda^2(0) + \sum_{r=1}^{\lambda} \frac{H_{\lambda-r}^2(0)}{2^{\lambda-r}(\lambda-r)!} \right\},$$

Note that initial integral (B.9) equals to $G_\lambda^{2\beta}$ at $\beta = \ell + \mu$. Formula (B.11) can be rewritten as

$$G_\lambda^{2\beta} = \left(\beta + \lambda - \frac{1}{2}\right) G_\lambda^{2(\beta-1)} - \lambda G_{\lambda-1}^{2(\beta-1)}, \quad \beta \in \mathbb{N}. \tag{B.15}$$

Substituting expression (B.14) into the recurrent formula (B.15) it is possible to calculate the values of $G_\lambda^{2\beta}$, $\beta = \ell + \mu$. Thus

$$J_\lambda^{\ell,s}(x) = 2(m\hbar\omega)^{\ell+\frac{1}{2}} \sum_{\mu=0}^{s} C_s^\mu G_\lambda^{2(\ell+\mu)} (\kappa x)^{2(s-\mu)}. \tag{B.16}$$

Finally, for integral $I_{n,k}^\ell$ (B.7), we obtain

$$I_{n,k}^\ell = A_{n,k} 2^{|n-k|} |n-k| (m\hbar\omega)^{\ell+\frac{1}{2}} \sum_{\lambda=0}^{\min(n,k)} \sum_{s=0}^{[|n-k|/2]} (-1)^s \frac{(|n-k|-s-1)!}{4^s s! (|n-k|-2s)!} \times$$
$$\times \sum_{\mu=0}^{s} C_s^\mu G_\lambda^{2(\ell+\mu)} (\kappa x)^{|n-k|-2\mu} L_{\min(n,k)-\lambda}^{(|n-k|-1)} (2\kappa^2 x^2) e^{-\kappa^2 x^2}. \tag{B.17}$$

If relation (A.8) $\frac{1}{s} C_{|n-k|-(s+1)}^{s-1} = \frac{(|n-k|-s-1)!}{s!(|n-k|-2s)!}$, is taken into consideration then expression (B.17) takes the form:

$$I_{n,k}^\ell = (-1)^{\min(n,k)} \frac{(m\hbar\omega)^{\ell+\frac{1}{2}}}{\pi\hbar} 2^{\frac{3}{2}|n-k|} \sqrt{\frac{\min(n,k)!}{\max(n,k)!}} |n-k| \times$$
$$\times \sum_{\lambda=0}^{\min(n,k)} \sum_{s=0}^{[|n-k|/2]} \sum_{\mu=0}^{s} \frac{(-1)^s}{4^s s} C_{|n-k|-s-1}^{s-1} C_s^\mu G_\lambda^{2(\ell+\mu)} \bar{x}^{|n-k|-2\mu} L_{\min(n,k)-\lambda}^{(|n-k|-1)} (2\bar{x}^2) e^{-\bar{x}^2}. \tag{B.18}$$

In the case when $n = k$ it is impossible to use formula (B.6), therefore integral $I_{n,n}^\ell(x)$ has the form (A.11). Theorem 2 is proved.

**Appendix C**

*Proof of Theorem 3*

Let us consider integral (1.12):

$$R_s^{(1)}(p)\langle x^r \rangle_s(p) = \sum_{n,k=0}^{+\infty} \rho_{n,k}^{(s)} \int_{-\infty}^{+\infty} x^r w_{n,k}(x,p)dx =$$

$$= \frac{1}{\pi\hbar} \sum_{n,k=0}^{+\infty} \rho_{n,k}^{(s)} \int_{-\infty}^{+\infty} x^r e^{-2\varepsilon} \Upsilon_{n,k}\left(\sqrt{2\varepsilon}\right) \cos\left[(n-k)\varphi\right] dx + \tag{C.1}$$

$$+ \frac{i}{\pi\hbar} \sum_{n,k=0}^{+\infty} \rho_{n,k}^{(s)} \int_{-\infty}^{+\infty} x^r e^{-2\varepsilon} \Upsilon_{n,k}\left(\sqrt{2\varepsilon}\right) \sin\left[(n-k)\varphi\right] dx.$$

Since, according to the conditions of the theorem $\rho_{n,k}^{(s)} = \rho_{k,n}^{(s)}$, then due to the oddness of function $\sin\left[(n-k)\varphi\right] = -\sin\left[(k-n)\varphi\right]$ when summing over indices $n,k$ the imaginary component of expression (C.1) will vanish. Let us consider the real component of expression (C.1):

$$R_s^{(1)}\langle x^r \rangle_{s,p} = \sum_{n,k=0}^{+\infty} \rho_{n,k}^{(s)} J_{n,k}^r(p),$$

$$J_{n,k}^r(p) = \frac{1}{\pi\hbar} \int_{-\infty}^{+\infty} x^r e^{-2\varepsilon} \Upsilon_{n,k}\left(\sqrt{2\varepsilon}\right) \cos\left[(n-k)\varphi\right] dx. \tag{C.2}$$

When integrating over variable $x \in (-\infty, +\infty)$, angle $\varphi$ changes from $\pi$ to $0$. The value of $x=0$ corresponds to angle $\varphi = \frac{\pi}{2}$. On the indicated interval function $\cos\left[(n-k)\varphi\right]$ is even when $|n-k| = 2j$, and it is odd when $|n-k| = 2j+1$. Function $\Upsilon_{n,k}\left(\sqrt{2\varepsilon}\right)$ is even in variable $x$. Consequently, integral (C.2) $J_{n,k}^{2\ell}(p) = 0$ when $|n-k| = 2j+1$ and $J_{n,k}^{2\ell+1}(p) = 0$ when $|n-k| = 2j$. Let us calculate integral (C.2).

$$J_{n,k}^r(p) = \frac{\sqrt{n!k!}}{\pi\hbar} \frac{|n-k|}{2} 2^{|n-k|+n+k} \sum_{\lambda=0}^{\min(n,k)} \sum_{s=0}^{[|n-k|/2]} \frac{(-1)^{\lambda+s}}{2^{2(\lambda+s)} \lambda!(k-\lambda)!(n-\lambda)! s!(|n-k|-2s)!} \times$$

$$\times \int_{-\infty}^{+\infty} x^r e^{-2\varepsilon} \varepsilon^{\frac{n+k}{2}-\lambda} \overline{x}^{|n-k|-2s} (2\varepsilon)^{-\frac{|n-k|-2s}{2}} dx = \frac{1}{\pi\hbar\kappa^{r+1}} \frac{|n-k|}{\sqrt{n!k!}} 2^{\frac{|n-k|}{2}+n+k-1} \times$$

$$\times \sum_{\lambda=0}^{\min(n,k)} \sum_{s=0}^{[|n-k|/2]} \frac{(-1)^{\lambda+s} \lambda!}{2^{2(\lambda+s)-s} s} C_k^\lambda C_n^\lambda C_{|n-k|-s-1}^{s-1} \times \int_{-\infty}^{+\infty} e^{-2\varepsilon} \varepsilon^{\frac{n+k-|n-k|}{2}+s-\lambda} \overline{x}^{|n-k|-2s+r} d\overline{x},$$

$$J_{n,k}^r(p) = \frac{1}{\pi\hbar\kappa^{r+1}} \frac{|n-k|}{\sqrt{n!k!}} 2^{\frac{|n-k|}{2}-\min(n,k)+n+k-1} \times$$

$$\times \sum_{\lambda=0}^{\min(n,k)} \sum_{s=0}^{[|n-k|/2]} \frac{(-1)^{\lambda+s} 2^{\lambda-s} \lambda!}{2^{2(\lambda+s)-s} s} C_k^\lambda C_n^\lambda C_{|n-k|-s-1}^{s-1} e^{-\overline{p}^2} \int_{-\infty}^{+\infty} e^{-\overline{x}^2} \left(\overline{p}^2 + \overline{x}^2\right)^{\min(n,k)+s-\lambda} \overline{x}^{|n-k|-2s+r} d\overline{x},$$

$$J_{n,k}^r(p) = \frac{|n-k|}{2\pi\hbar\kappa^{r+1}} \sqrt{\frac{2^{3\max(n,k)-\min(n,k)}}{n!k!}} \sum_{\lambda=0}^{\min(n,k)} \sum_{s=0}^{[|n-k|/2]} \sum_{\mu=0}^{\min(n,k)+s-\lambda} \frac{(-1)^{\lambda+s} \lambda!}{2^{\lambda+2s} s} \times$$

$$\times C_{\min(n,k)+s-\lambda}^\mu C_k^\lambda C_n^\lambda C_{|n-k|-s-1}^{s-1} \overline{p}^{2(\min(n,k)+s-\lambda-\mu)} e^{-\overline{p}^2} \int_{-\infty}^{+\infty} e^{-\overline{x}^2} \overline{x}^{|n-k|+r+2(\mu-s)} d\overline{x}. \tag{C.3}$$

Integral from expression (C.3) can be represented as:

$$\int_{-\infty}^{+\infty} e^{-\bar{x}^2} \bar{x}^{|n-k|+r+2(\mu-s)} d\bar{x} = \begin{cases} 0, \text{ если } |n-k|+r = 2\nu+1, \\ \dfrac{|2(\nu+\mu-s)-1|!!}{2^{\nu+\mu-s}} \sqrt{\pi}, \text{ если } |n-k|+r = 2\nu. \end{cases} \quad (C.4)$$

Considering expression (C.4), integral (C.3) takes the form

$$J_{n,k}^{r}(p) = \begin{cases} 0, \text{ если } |n-k|+r = 2\nu+1, \\ \dfrac{|n-k|}{2\hbar\kappa^{r+1}} \sqrt{\dfrac{2^{3\max(n,k)-\min(n,k)}}{\pi n! k!}} \sum_{\lambda=0}^{\min(n,k)} \sum_{s=0}^{[|n-k|/2]} \sum_{\mu=0}^{\min(n,k)+s-\lambda} \dfrac{(-1)^{\lambda+s} \lambda!}{2^{\lambda+s+\mu+\nu} s} |2(\nu+\mu-s)-1|!! \times \\ \times C_{\min(n,k)+s-\lambda}^{\mu} C_{k}^{\lambda} C_{n}^{\lambda} C_{|n-k|-s-1}^{s-1} \bar{p}^{2(\min(n,k)+s-\lambda-\mu)} e^{-\bar{p}^2}, \text{ если } |n-k|+r = 2\nu. \end{cases} \quad (C.5)$$

If $n = k$, integral $J_{n,n}^{2\ell+1}(p) = 0$, therefore it is sufficient to calculate only $J_{n,n}^{2\ell}(p)$:

$$J_{n,n}^{2\ell}(p) = \dfrac{2}{\pi\hbar} \int_0^{+\infty} x^{2\ell} e^{-2\varepsilon} \Upsilon_{n,n}(\sqrt{2\varepsilon}) dx = \dfrac{n!}{\pi\hbar} 2^{2n+1} \sum_{\lambda=0}^{n} \dfrac{(-1)^{\lambda}}{2^{2\lambda} \lambda! [(n-\lambda)!]^2} \int_0^{+\infty} x^{2\ell} e^{-2\varepsilon} \varepsilon^{n-\lambda} dx =$$

$$= \dfrac{2^{n+1} n!}{\kappa^{2\ell+1} \pi\hbar} \sum_{\lambda=0}^{n} \dfrac{(-1)^{\lambda} e^{-\bar{p}^2}}{2^{\lambda} \lambda! [(n-\lambda)!]^2} \int_0^{+\infty} \bar{x}^{2\ell} e^{-\bar{x}^2} (\bar{p}^2 + \bar{x}^2)^{n-\lambda} d\bar{x} = \dfrac{2^{n+1} n!}{\kappa^{2\ell+1} \pi\hbar} \sum_{\lambda=0}^{n} \dfrac{(-1)^{\lambda} e^{-\bar{p}^2}}{2^{\lambda} \lambda! [(n-\lambda)!]^2} \times$$

$$\times \sum_{\mu=0}^{n-\lambda} C_{n-\lambda}^{\mu} \bar{p}^{2(n-\lambda-\mu)} \int_0^{+\infty} \bar{x}^{2(\ell+\mu)} e^{-\bar{x}^2} d\bar{x},$$

$$J_{n,n}^{2\ell}(p) = \dfrac{2^{n} n!}{\kappa^{2\ell+1} \hbar\sqrt{\pi}} \sum_{\lambda=0}^{n} \sum_{\mu=0}^{n-\lambda} \dfrac{(-1)^{\lambda} |2(\ell+\mu)-1|!!}{2^{\lambda+\ell+\mu} \lambda! [(n-\lambda)!]^2} C_{n-\lambda}^{\mu} \bar{p}^{2(n-\lambda-\mu)} e^{-\bar{p}^2}. \quad (C.6)$$

Taking into consideration that $R_s^{(1)}(p) = \int_{-\infty}^{+\infty} W(x,p) dx = \sum_{n,k=0}^{+\infty} \rho_{n,k} J_{n,k}^{0}(p)$ and expressions (C.2), (C.5)-(C.6), we arrive at the validity of (1.13) and (1.14). Theorem 3 is proved.